\def\ga{\mathrel{\mathchoice {\vcenter{\offinterlineskip\halign{\hfil
$\displaystyle##$\hfil\cr>\cr\sim\cr}}}
{\vcenter{\offinterlineskip\halign{\hfil$\textstyle##$\hfil\cr
>\cr\sim\cr}}}
{\vcenter{\offinterlineskip\halign{\hfil$\scriptstyle##$\hfil\cr
>\cr\sim\cr}}}
{\vcenter{\offinterlineskip\halign{\hfil$\scriptscriptstyle##$\hfil\cr
>\cr\sim\cr}}}}}

\def\la{\mathrel{\mathchoice {\vcenter{\offinterlineskip\halign{\hfil
$\displaystyle##$\hfil\cr<\cr\sim\cr}}}
{\vcenter{\offinterlineskip\halign{\hfil$\textstyle##$\hfil\cr
<\cr\sim\cr}}}
{\vcenter{\offinterlineskip\halign{\hfil$\scriptstyle##$\hfil\cr
<\cr\sim\cr}}}
{\vcenter{\offinterlineskip\halign{\hfil$\scriptscriptstyle##$\hfil\cr
<\cr\sim\cr}}}}}
\documentclass{iaus}
\usepackage{graphicx}
\usepackage{natbib}

\title[Fullerenes in space] 
{Fullerenes in circumstellar and interstellar environments}

\author[J. Cami et al.]   
{Jan Cami$^{1,2}$ 
 \and Jeronimo Bernard-Salas$^{3,4}$ \and Els Peeters$^{1,2}$ \and Sarah
 E. Malek$^1$}

\affiliation{$^1$Department of Physics \& Astronomy, The University of
  Western Ontario, \\ London, ON N6A 3K7, Canada \\ email: {\tt
    jcami@uwo.ca} \\[\affilskip]
$^2$SETI Institute, 189 Bernardo Ave, Mountain View, CA 94043, USA\\[\affilskip]
$^3$Institut d'Astrophysique Spatiale, CNRS/Universit\'{e} Paris-Sud
11, 91405 Orsay, France\\[\affilskip]
$^4$Cornell University, 222 Space Sciences Bld., Ithaca, NY 14853, USA
}

\pubyear{2010}
\volume{280}  
\pagerange{119--126}
\jname{The Molecular Universe}
\editors{A.C. Editor, B.D. Editor \& C.E. Editor, eds.}
\begin{document}

\maketitle

\begin{abstract}
  We recently identified several emission bands in the Spitzer-IRS
  spectrum of the unusual planetary nebula Tc~1 with the infrared
  active vibrational modes of the neutral fullerene species C$_{60}$
  and C$_{70}$. Since then, the fullerene bands have been detected in
  a variety of sources representing circumstellar and interstellar
  environments. Abundance estimates suggest that C$_{60}$ represents
  $\sim$0.1\%--1.5\% of the available carbon in those sources. The
  observed relative band intensities in various sources are not fully
  compatible with single-photon heating and fluorescent cooling, and
  are better reproduced by a thermal distribution at least in some
  sources. The observational data suggests that fullerenes form in the
  circumstellar environments of evolved stars, and survive in the
  interstellar medium. Precisely how they form is still a matter of
  debate.  \keywords{astrochemistry, circumstellar matter,
    ISM:molecules, infrared:ISM}
\end{abstract}

\firstsection 

\section{Introduction}

\begin{figure}
\centering
\resizebox{\hsize}{!}{%
\includegraphics{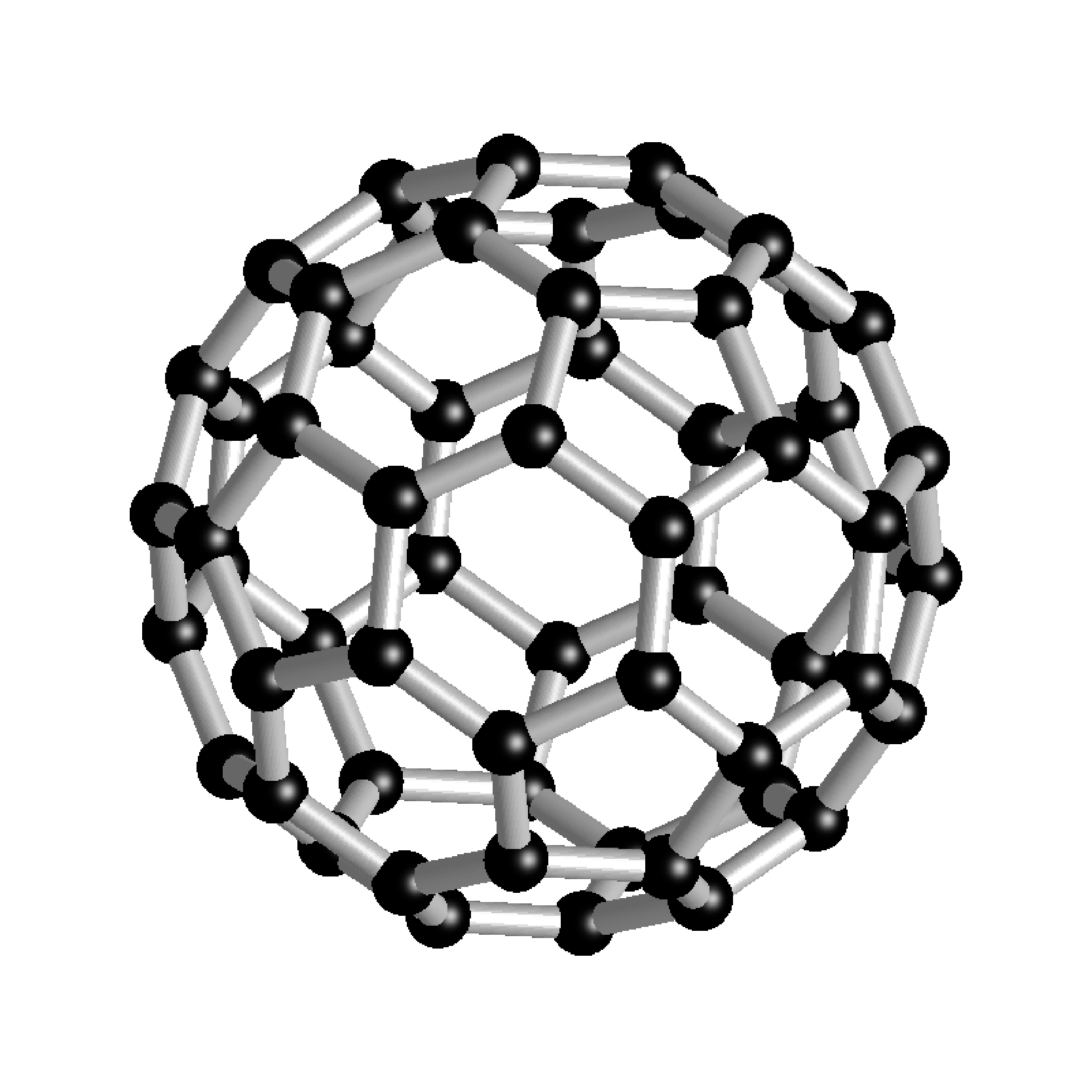}%
\includegraphics{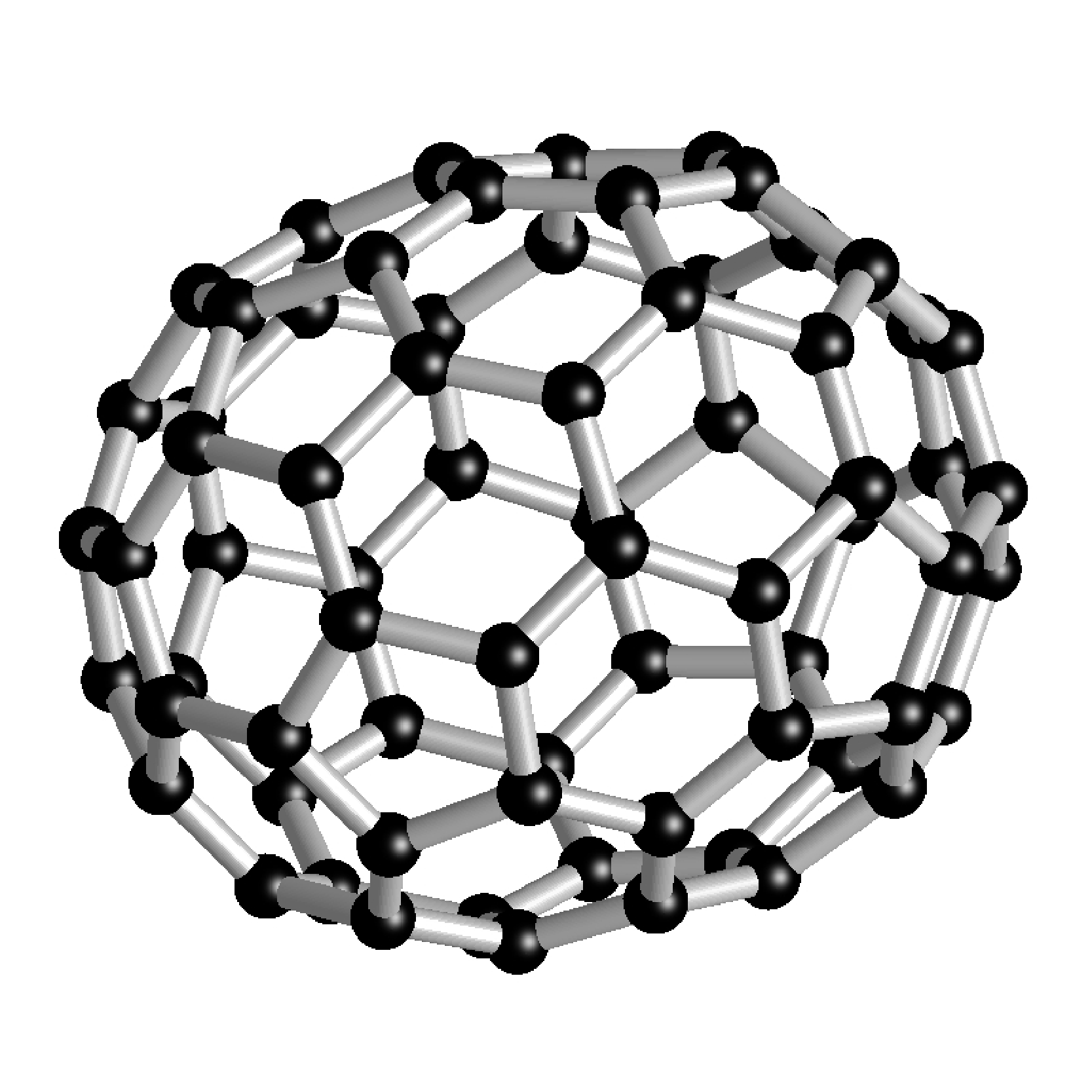}}
\caption{\label{Fig:structure}A representation of the molecular
  structure of C$_{60}$ (left) and C$_{70}$ (right). }
\end{figure}

Fullerenes are large molecules made of carbon hexagons and pentagons
that are organized in the shape of a hollow sphere or ellipsoid. The
best-known member of the class is the archetypal so-called
``buckminsterfullerene'' C$_{60}$ whose structure is the same as a
soccer ball (see Fig.~\ref{Fig:structure}).  Fullerenes were
discovered in a series of laboratory experiments that simulated the
circumstellar environment of carbon stars and that were aimed at
understanding the formation of long carbon chains in such outflows
\citep{1985Natur.318..162K}. Their unique properties has made them the
topic of much research in very diverse fields on Earth -- from
superconductivity over nanotechnology to targeted drug delivery.

As soon as they were discovered on earth, it was suggested that
fullerenes could also form in carbon star outflows and be injected
into the interstellar medium (ISM). Since they are amongst the most
stable carbon species known, they would then be ideally suited to
survive the harsh conditions in the ISM.

Dedicated searches for the electronic bands of neutral C$_{60}$ in the
ISM and in circumstellar shells were negative
\citep{1989A&A...213..291S,1989MNRAS.240P..41S,Herbig:C60}.
\citet{Foing:C60_1} on the other hand reported the detection of two
diffuse interstellar bands (DIBs) whose wavelengths are consistent
with transitions of the C$_{60}^+$ cation as measured in solid
matrices \citep{1993CPL...211..227F}; more recently, several more DIBs
have been found that could be due to the same cation
\citep{2009ApJ...700.1988M}. However, since the technique of matrix
isolation spectroscopy introduces unpredictable shifts in the
frequencies of the absorption bands, this promising case awaits 
confirmation from a comparison to a cold gas phase spectrum which is
not yet available. 

Fullerenes can also be detected through their vibrational modes at
infrared (IR) wavelengths (see Sect.~\ref{Sect:SpecProp}), and several
efforts have focused on these spectral signatures
\citep{Moutou:C60,1995AJ....109.2096C}. Again, no conclusive evidence
was found for the presence of C$_{60}$ in space. More recently,
\citet{2007ApJ...659.1338S} analyzed the 15--20 $\mu$m spectrum of the
reflection nebula NGC~7023 obtained with the Infrared Spectrograph
\citep[IRS,][]{Houck:IRS} on board the Spitzer Space Telescope
\citep{Werner:spitzer}. They suggested that C$_{60}$ could be one of
the possible carriers for the spectral features in this source at 17.4
and 18.9 $\mu$m.

\citet{Cami:C60-Science} presented the unusual Spitzer-IRS spectrum of
the planetary nebula (PN) Tc~1, and identified several emission
features with the vibrational modes of the fullerene species C$_{60}$
and C$_{70}$. Since then, fullerenes have been confirmed in many more
sources of different nature, indicating that these species may be
quite widespread in the Universe. A lively discussion has ensued
especially over their excitation properties and formation mechanism. 

Here, we briefly review some of the relevant properties of C$_{60}$
and C$_{70}$ in Sect.~\ref{Sect:SpecProp}. We present the case for
fullerenes in Tc~1 in detail, and summarize the other detections in
Sect.~\ref{Sect:detection}. Some aspects of the excitation mechanism
are discussed in Sect.~\ref{Sect:Excitation}, and the formation of
fullerenes in space in Sect.~\ref{Sect:Formation}.

\section{Spectroscopic properties of C$_{60}$ and C$_{70}$}
\label{Sect:SpecProp}

A representation of a C$_{60}$ molecule is shown in
Fig.~\ref{Fig:structure}. The 60 carbon atoms are arranged in the
shape of a truncated icosahedron ($I_h$ symmetry): a hollow sphere
that is made of 20 hexagons and 12 pentagons. C$_{60}$ has 174
vibrational modes, but owing to its symmetry, many modes are
degenerate and only 4 distinct modes are IR active: the
$T_{1u}(1)$ mode at 18.9 $\mu$m; the $T_{1u}(2)$ mode at 17.4 $\mu$m;
the $T_{1u}(3)$ mode at 8.5 $\mu$m and the $T_{1u}(4)$ mode at 7.0
$\mu$m. Laboratory experiments have measured the frequencies of these
transitions in the pure, solid state
\citep{1990Natur.347..354K,vonCzarnowski1995321} as well as in the
gas-phase \citep{Frum:C60,Nemes:C60C70_temp}. In those experiments,
the measured bands are fairly broad and symmetric, and thus well
characterized by a central wavelength and a width. The differences
between the measured frequencies primarily reflect the effect of
temperature: a higher temperature shifts the bands to lower
frequencies and causes broadening \citep{Nemes:C60C70_temp}.

The intrinsic strength of the C$_{60}$ bands is less well known, and
the literature offers quite different values for the absorption
coefficients (even relative to one another), obtained either from
laboratory experiments or theoretical
calculations. \citet{Cami:C60-Science} used data from
\citet{PhysRevB.47.14607,PhysRevB.53.13864,ISI:000087900400035}; for
the analysis of NGC~7023 by \citet{2010ApJ...722L..54S}, strengths
from \citet{Choi_etal_2000} were used. Also
\citet{2011MNRAS.413..213I} presented molar absorptivities derived
from laboratory experiments. 

Whereas the model for the C$_{60}$ geometry has traditionally been a
soccer ball, the C$_{70}$ molecule is a prolate top more like a rugby
ball. The lower symmetry (D$_{5h}$) results in more IR active modes:
of the 204 fundamental vibrational modes, 31 are IR active, and
laboratory studies as well as theoretical DFT calculations are
available
\citep{vonCzarnowski1995321,ISI:000074726600006,ISI:000174398500021}. Most
of the C$_{70}$ IR active bands are weak, and furthermore blend with
the C$_{60}$ bands; but several isolated bands of medium strength are
at 14.8, 21.8, 12.6 and 15.6 $\mu$m.

\section{Fullerenes in astrophysical environments}
\label{Sect:detection}

\subsection{The planetary nebula Tc~1}

\citet{Cami:C60-Science} reported the identification of the IR
C$_{60}$ bands in the unusual Spitzer-IRS spectrum of the planetary
nebula (PN) Tc~1. The high quality of the spectrum and the lack of
contaminating features make this source the clearest case for the
presence of fullerenes in space. 

\begin{figure}
\centering
\resizebox{\hsize}{!}{\includegraphics{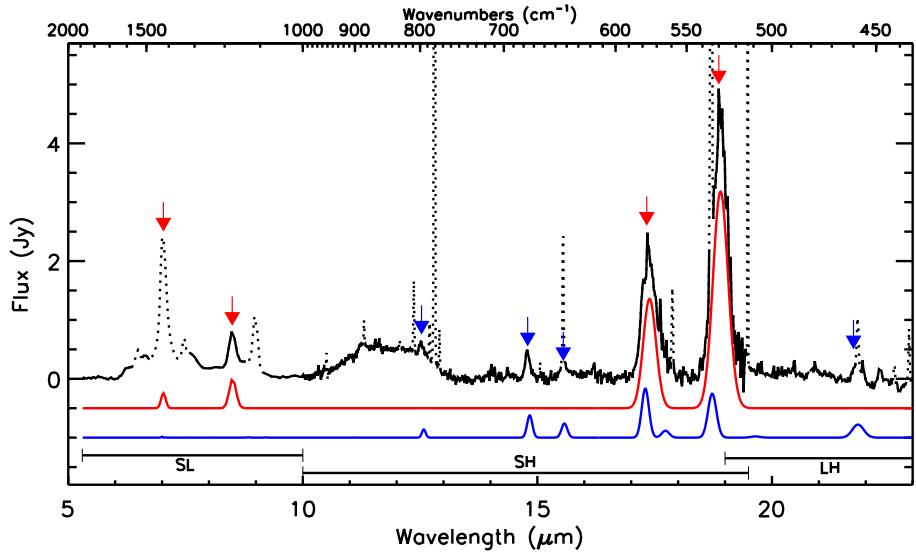}}
\caption{\label{Fig:Tc1spec}A composite IR spectrum of the planetary
  nebula Tc~1 as observed by Spitzer-IRS is shown on top. The spectrum
  is the combination of observations with the Short-Low module (SL,
  $\lambda / \Delta \lambda \sim 60-120$) and Short- and Long-High (SH
  and LH, $\lambda/\Delta\lambda \sim 600$) modules. A featureless
  dust continuum was subtracted from the spectrum. Known strong
  emission lines are masked and shown here as dotted lines. The curve
  below it is a thermal emission model for C$_{60}$; the bottom curve
  a similar model for C$_{70}$. Figure taken from
  \citet{Cami:C60-Science}.}
\end{figure}

Fig.~\ref{Fig:Tc1spec} shows the continuum-subtracted Spitzer-IRS
observations of the planetary nebula Tc~1. Apart from the strong
(typically forbidden) emission lines that are characteristic for PNe,
the spectrum is very unusual, and exhibits clear emission bands at the
wavelengths of all four fundamental modes of C$_{60}$\footnote{Note
  however that a significant fraction of the strong emission at 7.0
  $\mu$m is due to an [ArII] line.}. Weaker and narrower features are
present at the wavelengths of the strongest (unblended) C$_{70}$
bands. In addition to the fullerene bands, a broad underlying plateau
is evident between 6--9 $\mu$m, as is a broad emission feature at 11.5
$\mu$m that is generally attributed to SiC \citep{Speck:SiC}. There is
not much else left in the spectrum though, and thus little
contamination confuses the analysis of the fullerene spectral
features.

As mentioned in Sect.~\ref{Sect:SpecProp}, the experimentally obtained
central wavelengths and band widths for the C$_{60}$ bands depend on
the temperature, and thus a comparison with laboratory data requires
some estimate of a typical temperature for the fullerenes in
Tc~1. Surprisingly, the observed band strengths for both C$_{60}$ and
C$_{70}$ are consistent with a thermal population distribution over
the vibrational levels (see Sect.~\ref{Sect:Excitation}). For
C$_{60}$, we can thus derive an excitation temperature of 332~K; for
C$_{70}$ the derived temperature is 179~K. All other measurable
quantities are consistent with such a temperature. Indeed, the central
wavelengths of all C$_{60}$ emission bands in the spectrum of Tc~1
correspond especially well to experimental values measured at
$\sim$300~K (rather than values measured at e.g. 1000~K). Further
support comes from the band widths: laboratory spectra show
temperature-dependent widths (FWHM) from about 8 cm$^{-1}$ at 300~K to
13 cm$^{-1}$ at 1000~K \citep{Nemes:C60C70_temp} for both the 17.4 and
18.9 $\mu$m C$_{60}$ bands. For Tc~1, the measured band widths are 9.9
and 12.6 cm$^{-1}$ for the bands at 18.9 and 17.4 $\mu$m respectively,
but since both bands suffer from contamination by the C$_{70}$ bands,
their true band widths are necessarily smaller, and thus only
compatible with the lower temperature regime. The C$_{60}$ bands at
7.0 and 8.5 $\mu$m are not resolved and thus offer no such
information. Also the central wavelengths of the four isolated
C$_{70}$ features compare better to the frequencies measured at low
temperatures.

Summarizing, we can conclude that the spectrum of Tc~1 shows the clear
spectral fingerprint of the fullerene species C$_{60}$ and C$_{70}$,
and that all measurable quantities (wavelengths, widths and band
ratios) are consistent with laboratory experiments at temperatures of
a few hundred K. 

\subsection{Other Evolved Objects}

With the clean fullerene spectrum of Tc~1 as a template, the presence
of C$_{60}$ has been confirmed in the circumstellar environments of a
variety of evolved objects. \citet{Garcia-Hernandez:PN} found the
C$_{60}$ bands in several other carbon-rich PNe. In most cases, the
spectra are contaminated by PAH features, thus complicating the
analysis; a notable exception is the PN SMP~SMC~16 in the Small
Magellanic Cloud that looks quite similar to Tc~1. No conclusive
confirmation of the weaker C$_{70}$ features has been reported in
these objects though. At first, these detections suggested fullerenes
form in the PN phase. However, they are already present in at least
one proto-PN object \citep{ZhangKwok:proto-PNC60}, and the C$_{60}$
bands are also seen in the Spitzer-IRS spectra of a few post-AGB stars
\citep{Gielen:C60p-AGB}. 

In the search for fullerenes in space, special attention was given to
the R Coronae Borealis (RCB) stars, a class of hydrogen-poor and
helium-rich supergiants. Under typical conditions in circumstellar
environments, the presence of hydrogen inhibits the formation of
fullerenes (see Sect.~\ref{Sect:Formation}), and thus the
hydrogen-poor nature of the circumstellar environments of RCB stars
was considered ideal to form C$_{60}$. The first search for C$_{60}$
in such environments did not turn up any evidence for fullerenes
\citep{1995AJ....109.2096C}. More recently, C$_{60}$ has been detected
in two RCB stars that were studied with the Spitzer Space Telescope
\citep{Garcia-Hernandez:RcrB}. Surprisingly, those two objects
represent the most H-rich objects in a larger sample of RCB stars. In
another study, an additional detection was made in one more RCB star
\citep{2011arXiv1106.0563C}.

\subsection{Interstellar environments}

The spectral features of C$_{60}$ have also been detected in
interstellar environments -- in the two reflection nebulae NGC 7023
and NGC 2023 \citep{2010ApJ...722L..54S}, and also in the Orion Bar
\citep{2011MNRAS.410.1320R}. These sources are spatially resolved, and
\citet{2010ApJ...722L..54S} showed that in NGC~7023, the peak emission
of PAHs originates from a different location as the peak emission from
C$_{60}$ (see Fig.~\ref{Fig:NGC7023:C60}). This has some implications
for the formation and excitation of fullerenes (see
Sect.~\ref{Sect:Excitation} and \ref{Sect:Formation}).

\subsection{Abundances}

With the presence of fullerenes in space firmly established, an
important question to address is what fraction of the cosmic carbon
they represent, and different methods have been applied to estimate
the fullerene abundances. In the evolved stars, of the order of
0.06--1.5 \% of the available carbon is estimated to be in the form of
C$_{60}$
\citep{Cami:C60-Science,Garcia-Hernandez:PN,ZhangKwok:proto-PNC60}. For
NGC~7023 on the other hand, \citet{2010ApJ...722L..54S} find that
0.1--0.6\% of the interstellar carbon is in C$_{60}$. This latter is
compatible with earlier estimates of 0.3--0.9\% of the cosmic carbon
that was estimated to be in the form of C$_{60}^+$
\citep{Foing:C60_1}. 

It is interesting to see that the abundance estimates in the evolved
objects are comparable to the interstellar fullerene abundances. If
only a fraction of the carbon-rich evolved stars would produce
fullerenes, one would expect the interstellar abundances to be much
lower than the circumstellar abundances in individual objects. If the
derived abundances are realistic, this then implies that the formation
of fullerenes must be fairly common. In any case, the derived
abundances represent a significant fraction of the cosmic carbon for a
single molecular species.

\bigskip

The presence of C$_{60}$ in diverse evolved objects suggests that
fullerenes form in circumstellar outflows whenever conditions are
right, and the detection of fullerenes in interstellar environments is
then a testimony to the stability of these species that allows them to
survive the harsh conditions in the ISM. In very general terms, the
available observational data on the fullerenes thus confirm the early
suggestion by \citet{1985Natur.318..162K}: fullerenes form in
carbon-rich circumstellar environments, and survive and thrive in the
interstellar medium.

\section{Excitation mechanism}
\label{Sect:Excitation}

\begin{figure}
\centering
\resizebox{\hsize}{!}{\includegraphics{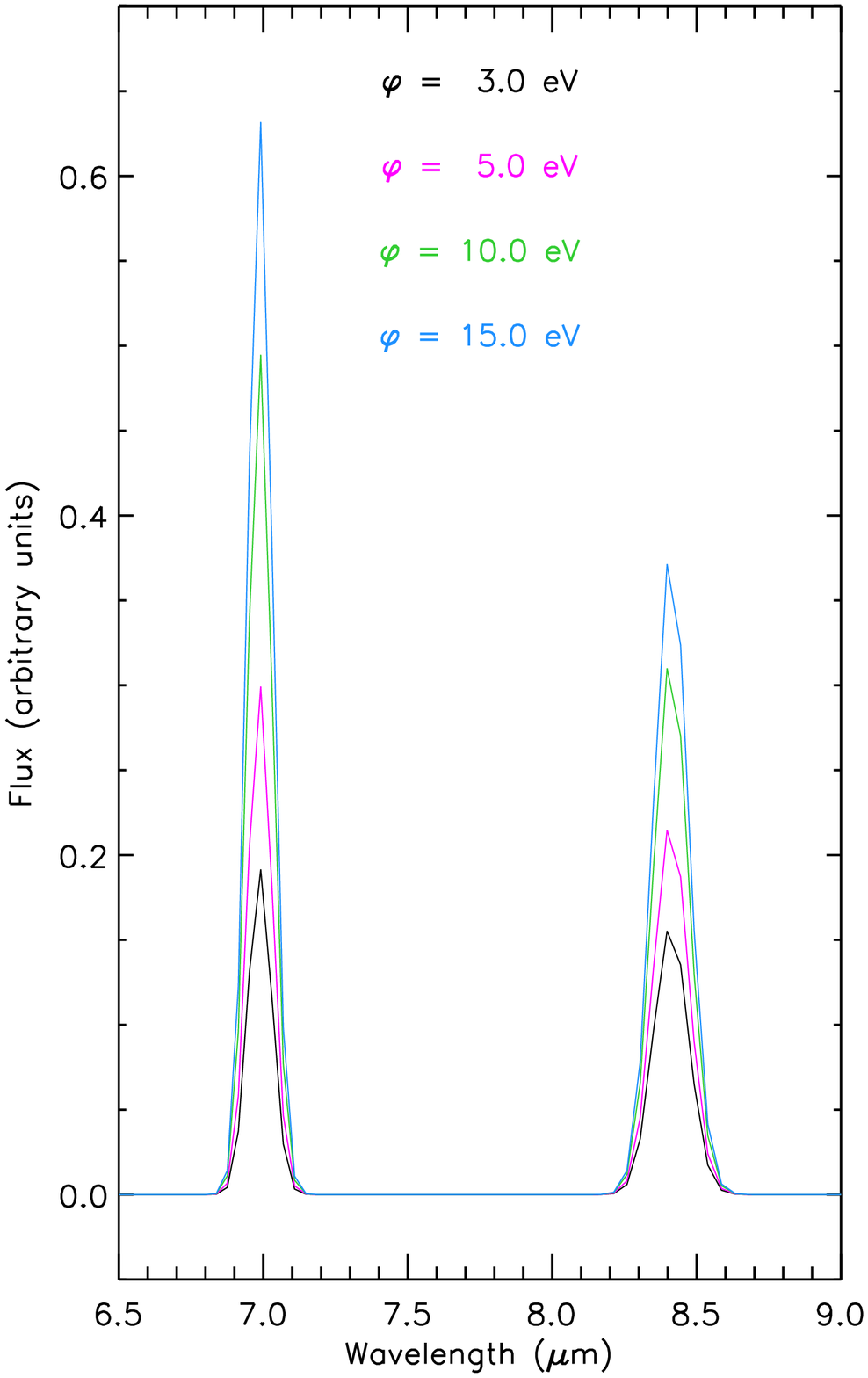}\includegraphics{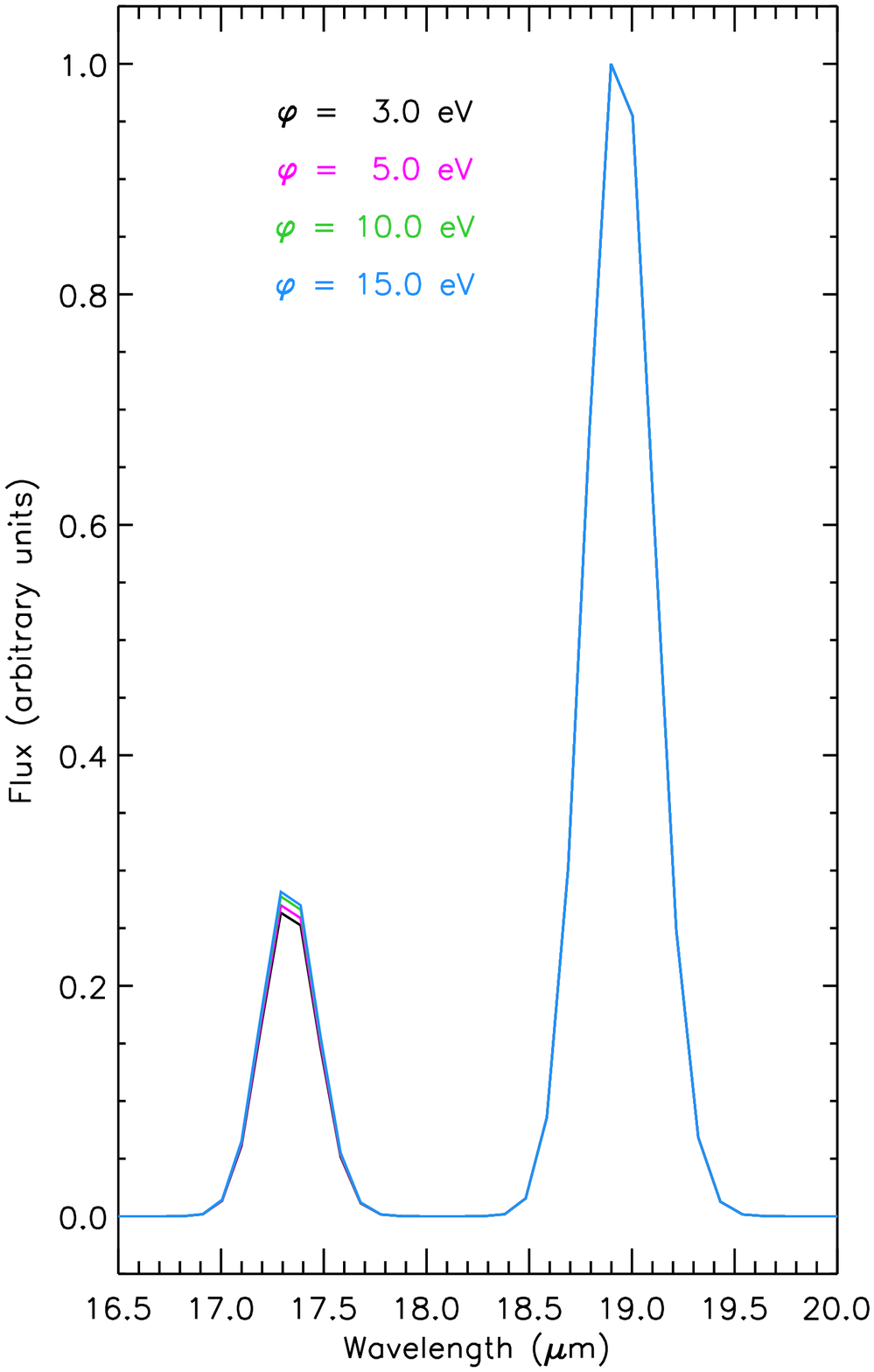}}
\caption{\label{Fig:cascademodels}A few illustrative examples of how
  the C$_{60}$ fluorescent emission spectrum changes for average
  photon energies of 3 eV, 5 eV, 10 eV and 15 eV (the lowest photon
  energies corresponding to the weakest emission). All spectra are
  normalized to the peak intensity of the 18.9 $\mu$m band. Widths
  were taken to be 10 cm$^{-1}$ for the 17.4 and 18.9 $\mu$m bands and
  20 cm$^{-1}$ for the bands at 7.0 and 8.5 $\mu$m -- consistent with
  the widths measured in the spectrum of Tc~1. }
\end{figure}

Having identified several emission bands with a specific molecular
carrier offers the prospect of using these spectral features as a
diagnostic tool to extract quantitative information about the physical
conditions that reign in the environment in which they
reside. Precisely what information can be obtained depends somewhat on
the molecules excitation mechanism at play. 

Large, free molecules such as fullerenes and PAHs are generally
thought to be excited by the absorption of a single photon (typically
in the UV range). Rapid internal conversion then leaves the molecule
in the electronic ground state, but in highly excited vibrational
states; and as the molecule relaxes, it emits IR photons. For all
relevant processes pertaining to the excitation \& cooling mechanism,
fullerenes have indeed very similar properties to PAHs \cite[see
e.g.][]{Tielens:book}. \citet{2010ApJ...722L..54S} analyzed the
C$_{60}$ in NGC~7023 using such an excitation model. 

The resulting emission spectrum of this fluorescent cooling process
can be calculated in various ways
\citep[e.g.][]{2001ApJ...560..261B,Joblin:MC}. The only free parameter
in these calculations is the average photon energy that is absorbed by
the molecules -- which is a function of the ambient radiation field
and the absorption cross section of the molecule in
question. Fig.~\ref{Fig:cascademodels} shows a few examples of how the
resulting IR spectra change following absorption of photons with
different energies. Note how there is little change in the ratio
between the 17.4 $\mu$m band and the 18.9 $\mu$m band for the energies
considered here. For typical PNe conditions ($\varphi \ga$ 7 eV)
however, the bands at 7.0 and 8.5 $\mu$m should be fairly strong, and
the band ratios $I_{7.0}/I_{18.9}$ and $I_{8.5}/I_{18.9}$ could in
principle be used as a sensitive measure of the average photon energy.

However, the numerical values for the band ratios depend on the
intrinsic band strengths used in the calculations; as mentioned in
Sect.~\ref{Sect:SpecProp}, these are not well known, and different
values are available in the literature. Using the intensities as given
by \citet{Choi_etal_2000} (as was done for the analysis of the
C$_{60}$ bands in the reflection nebula NGC~7023 presented by
\citet{2010ApJ...722L..54S}), the $I_{17.4}/I_{18.9}$ band ratio
should be roughly constant at about 0.28 for photon energies relevant
to PNe environments. The measured band ratio in Tc~1 however is 0.59;
when accounting for the contamination by C$_{70}$ the ratio is
somewhat lower, but can never be brought in agreement with the
expected ratio for fluorescent emission when adopting these intrinsic
strengths. The 8.5/18.9 $\mu$m band ratio on the other hand is too low
to be consistent with the same fluorescent cooling model.

One could argue that the adopted intensities for the bands are not
correct, but also the other observations of the C$_{60}$ bands pose
difficulties in the framework of fluorescence. For all detected
sources, the band ratios (including $I_{17.4}/I_{18.9}$) show
significant variations from source to source \citep[see
e.g.][]{Garcia-Hernandez:PN}. This is also evident from the different
excitation temperatures ($\sim$200--700~K) that are derived assuming a
thermal distribution over the vibrational bands. In many cases, the
bands are contaminated by PAH features, but it is not clear if that
alone can explain all the observed spectral variations. If true, these
variations (and especially the $I_{17.4}/I_{18.9}$ band ratio) are not
compatible with fluorescent cooling, irrespective of the intrinsic
strength values used. 

\bigskip

\begin{figure}
\resizebox{\hsize}{!}{%
\includegraphics{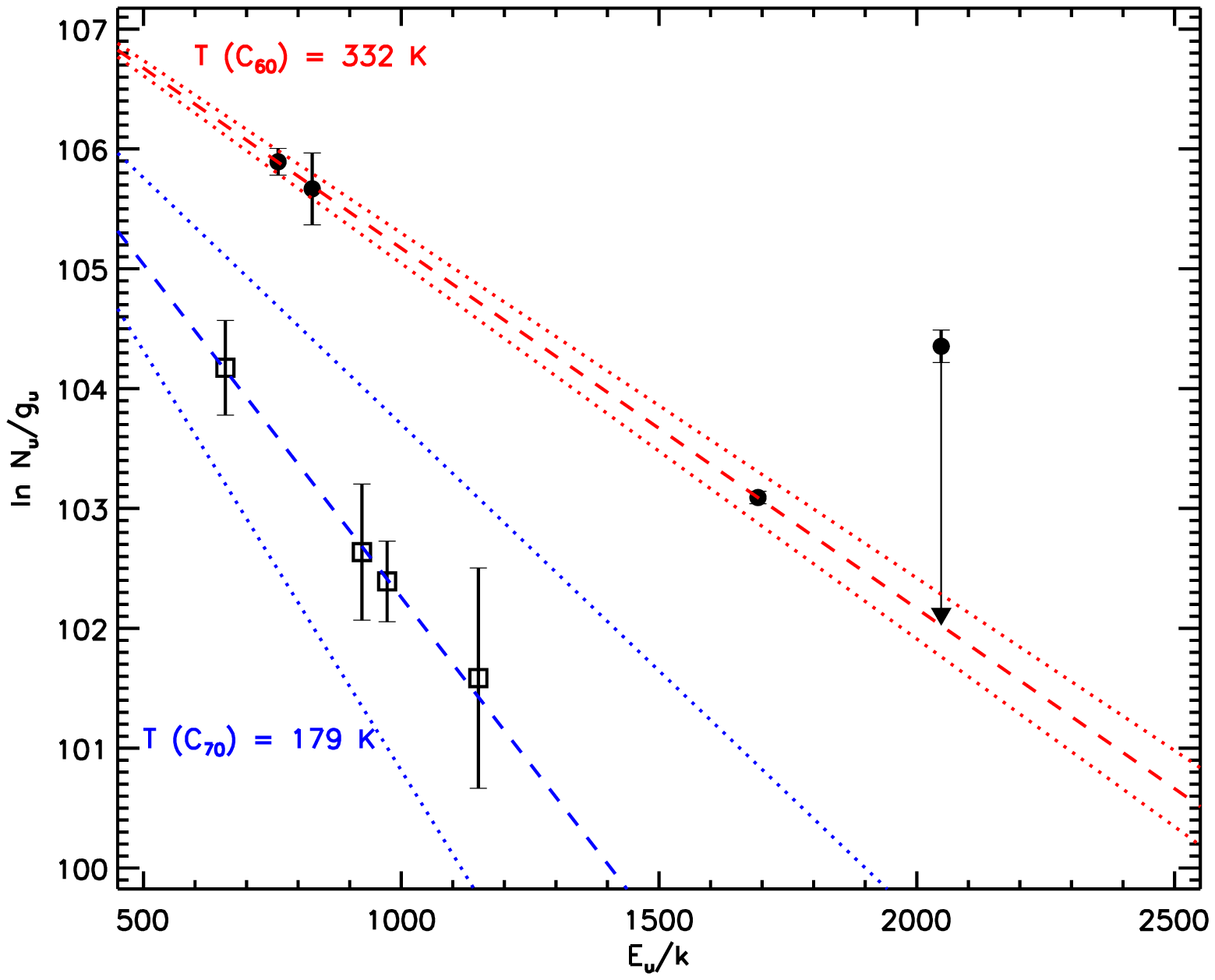}%
\includegraphics{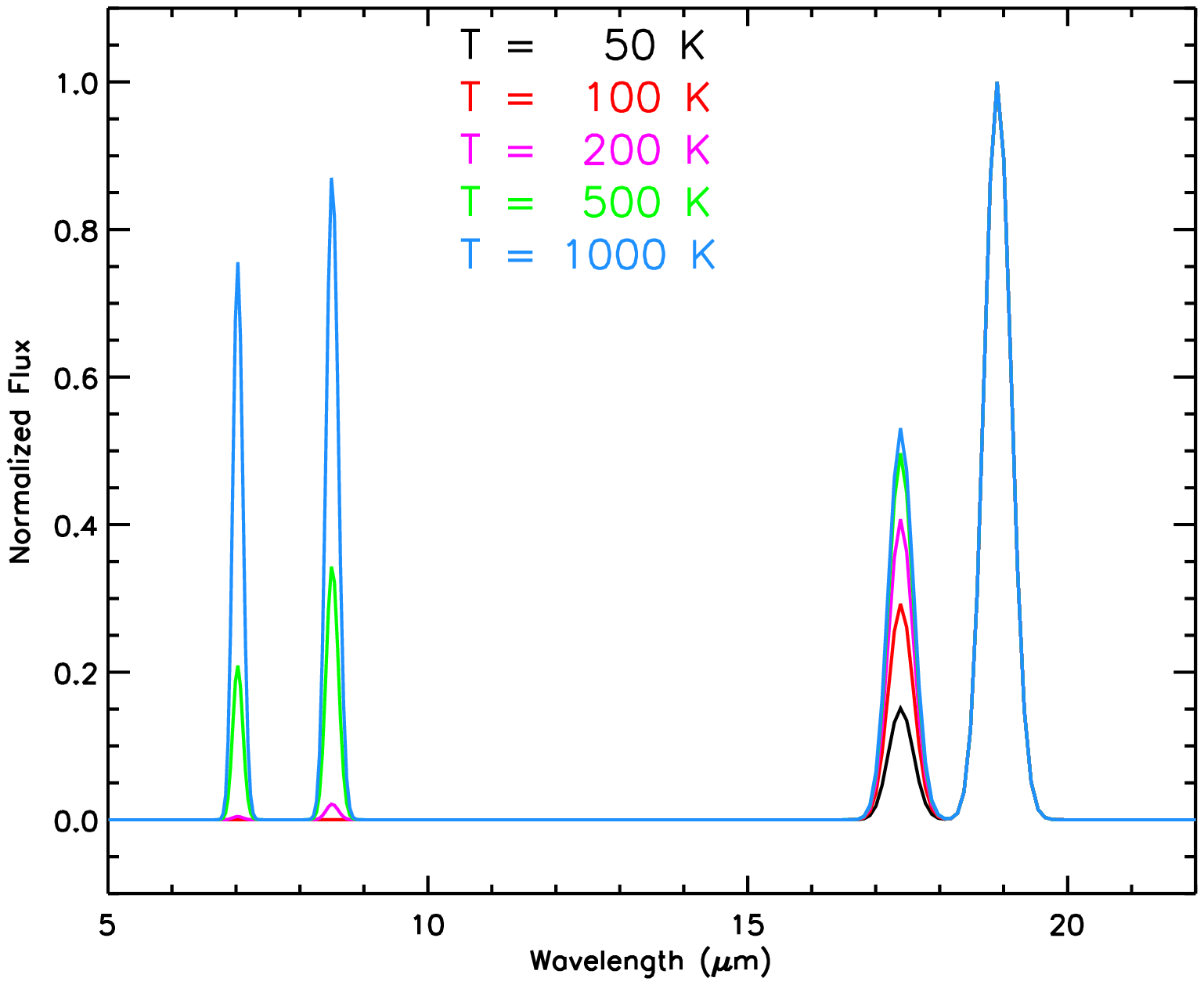}}
\caption{\label{Fig:thermalstuff}{\it (left)} Excitation diagram for
  the C$_{60}$ and C$_{70}$ bands in the Spitzer-IRS spectrum of Tc~1
  \citep[Fig. from][]{Cami:C60-Science}. {\it (right)} Illustrative
  examples of the spectral variations in the C$_{60}$ bands. The bands
  at 7.0 and 8.5 $\mu$m are virtually absent for excitation
  temperatures below $\sim$200~K, whereas the 17.4/18.9 $\mu$m band
  ratio shows large variations for temperatures up to $\sim$500~K.}
\end{figure}

The fairly clean fullerene spectrum of Tc~1 offers the possibility of
directly calculating the population distribution over the excited
vibrational states from the total emitted power in the bands; and this
can be done for both C$_{60}$ and C$_{70}$. Surprisingly, the
population distribution is fully consistent with a thermal
distribution (see Fig.~\ref{Fig:thermalstuff}) over the vibrational
states corresponding to an excitation temperature of 332~K for
C$_{60}$, and 179~K for C$_{70}$.

Thermal models show quite different spectral variations as a function
of temperature (Fig.~\ref{Fig:thermalstuff}) compared to the
fluorescent cooling models and can more easily accommodate some of the
observed changes in the $I_{17.4}/I_{18.9}$ band ratios. The cases
where the 7.0 and 8.5 $\mu$m band are missing would then correspond to
excitation temperatures $\la 200~K$, whereas an unacceptably low
photon energy would be required in the fluorescence model. 

It is not easy to understand how free C$_{60}$ molecules in
circumstellar or interstellar environments could produce an emission
spectrum that is consistent with thermal excitation rather than with
stochastic heating and fluorescent cooling. Since gas densities are
typically orders of magnitude lower than the critical density, it is
unlikely that gas collisions could keep the fullerenes thermalized. It
was thus proposed that the fullerene molecules are not free, but in
direct contact with the surface of dust grains, or in the solid state
themselves \citet{Cami:C60-Science}. In such a case, the fullerenes
would be in equilibrium with the dust grains, which themselves are in
equilibrium with the ambient radiation field. However, the derived
excitation temperatures are hard to understand in this
scenario. Further observations and analyses will be required to settle
the excitation mechanism conclusively.

\section{The formation of astrophysical fullerenes}
\label{Sect:Formation}

From the fairly large number of evolved objects in which the C$_{60}$
bands have been detected, it seems clear that these species do indeed
form in the circumstellar environments of evolved stars. Precisely how
this happens is far less clear though, and there has been some
discussion about the precise pathways to fullerene
formation\footnote{\citet{1995IJMSI.149..321B} additionally discuss a
  possible formation route for fullerenes under interstellar
  conditions.}.

In this context, it is worthwhile to consider what we know from
laboratory experiments. In the original experiments by
\citet{1985Natur.318..162K}, fullerenes self-assemble from much
smaller carbon fragments by collisions. At very low gas densities, a
broad distribution of cluster sizes is observed; but as the gas
density is increased, collisions become more numerous, and only the
most stable species (C$_{60}$) survives. If in carbon-rich
circumstellar environments, collisional processes (occurring under the
similar conditions) are driving the chemistry, fullerenes could form
in the same way. The efficiency of the fullerene formation process is
then a function of the density. However, an important condition in
this formation process is the absence of hydrogen. It has long been
realized that the presence of H will inhibit the formation of
fullerenes \citep[see e.g.][]{1988Natur.331..328K}. Indeed, laboratory
experiments show that the mere presence of hydrogen (as little as
0.4\%) in such conditions already greatly reduces fullerene
formation. At a hydrogen fraction of 10\%, no fullerenes are formed
anymore, and the experiments result in large amounts of PAHs instead
\citep{1993GeCoA..57..933D,1995JMatR..10.1977W}. 

Based on the Spitzer-IRS spectrum of Tc~1 alone, it would seem that
its circumstellar environment is fairly analogous to this laboratory
setup. Indeed, the available material is clearly carbon-rich, while at
the same time, the fullerene-rich IR spectrum does not show much
evidence for the presence of PAHs nor any other H-containing
species. It is thus conceivable that the region where the fullerenes
are located is hydrogen-poor. However, it is clear that H is present
at large in the nebula \citep[see
e.g.][]{2009AstL...35..518M,2008ApJ...677.1100W}, and even in the
photosphere of the central star \citep{1988A&A...190..113M}. Thus, if
the formation of fullerenes in Tc~1 involves collisional growth of
carbon clusters under similar conditions as in the laboratory
experiments, there must have been H-poor regions in the nebula to
allow the fullerenes to form.

\citet{Garcia-Hernandez:PN} argue against a formation scenario in
which a H-poor environment is required. They note that in their PNe
data, the Spitzer-IRS spectra show the simultaneous presence of the
C$_{60}$ features and PAHs. From this, they conclude that both species
co-exist, and that the fullerenes form at the same time and place as
PAHs, in a H-rich environment. They propose photo-processing of solid
Hydrogenated Amorphous Carbon (HAC) as the most plausible fullerene
formation route in PNe environments. Support for that assertion is
found in experimental results that show how such processing produces
PAHs and fullerenes at the same time
\citep{1997ApJ...489L.193S}. Moreover, all C$_{60}$ PNe furthermore
show the so-called 30 $\mu$m feature that is generally attributed to
MgS \citep[see e.g.][]{Hony:MgS}, but that also has been linked to
HACs \citep{2001ApJ...558L.129G}. They then interpret the PAH-poor and
fullerene-dominated sources Tc~1 and SMP~SMC~16 as environments where
the more hardy C$_{60}$ molecules have survived whereas the PAHs have
been destroyed.

It is clear that -- in the context of astrophysical fullerenes -- this
scenario presents several difficulties as well. First, if HAC
photo-processing is the general mechanism to produce PAHs and
fullerenes in PNe, and if the 30 $\mu$m feature is indeed related to
HACs, one would expect all PNe that exhibit the 30 $\mu$m feature to
also exhibit PAH and fullerene bands. Second, it seems unlikely that
UV photo-processing is a viable mechanism to explain the formation of
C$_{60}$ in evolved stars with much lower UV irradiation than is the
case in the PNe -- such as the proto-PN IRAS 01005+7910
\citep{ZhangKwok:proto-PNC60} but especially the two post-AGB stars
presented by \citet{Gielen:C60p-AGB} with effective temperatures of
the order of only 6,000~K. Finally, if Tc~1 and SMP~SMC~16 represent
environments where the fullerenes have survived and PAHs have not, it
is hard to explain the numerous PNe that show clear PAH features but
no C$_{60}$ bands.

It is important to realize that the mixture of carbonaceous species
that results from HAC photo-processing in the laboratory experiments is
quite different from what is observed in the C$_{60}$ PNe. The mass
spectra reported by \citet{1997ApJ...489L.193S} do indeed show the
presence of fullerenes, but fullerenes certainly do not dominate the
mass distribution. In all experiments, the photo-processing products
include many more (dehydrogenated) PAHs, and low-mass molecules (and
molecular fragments) containing H. Thus, to explain objects such as
Tc~1 and SMP~SMC~16 where abundant C$_{60}$ is seen but no PAHs nor
small species are present, significant {\em additional} processing of
these sputtered species would be required to destroy the smaller
molecules and the PAHs. UV radiation could conceivably dissociate some
of the smaller molecules, but it is not clear how PAHs could be
destroyed by radiation in the low-excitation environments represented
by the C$_{60}$ PNe. If, on the other hand, collisions are involved,
the carbon clusters are processed under similar conditions as
described above -- and either result in fullerenes dominating the
distribution (in the absence of H), or in the formation of PAHs.

\begin{figure}
\centering
\resizebox{7cm}{!}{\includegraphics{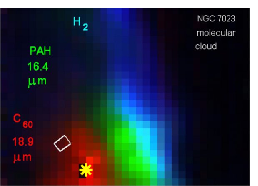}}
\caption{\label{Fig:NGC7023:C60}The spatial distribution of molecular
  hydrogen, PAHs and C$_{60}$ in the reflection nebula NGC 7023
  \citep[Fig. from][]{2010ApJ...722L..54S} }
\end{figure}

Finally, it should be noted that the simultaneous presence of the
C$_{60}$ bands and the PAH features in the spectra of some PNe does
not necessarily imply that they are co-located -- merely that they are
both located in the Spitzer beam. In fact, there is clear
observational evidence in another object that PAHs and fullerenes are
{\em not} co-located. In Fig.~\ref{Fig:NGC7023:C60}, we have
reproduced the spatial distribution of molecular hydrogen, PAHs and
fullerenes in the reflection nebula NGC~7023 \citep[][their
Fig.~3]{2010ApJ...722L..54S}. The C$_{60}$ emission peaks on the
central star, whereas the PAHs are clearly located further away from
the star. Thus, although the spectra of the entire nebula show both
PAH and fullerene features, both components show clearly different
spatial distributions -- no PAH emission is found where the C$_{60}$
peaks. Such a distribution could well be the case for (some of) the
PNe as well. \\

The two processes discussed above are not the only ways to form
circumstellar fullerenes. An important third route involves
high-temperature processes, and allows the formation of fullerenes in
H-rich environments. Laboratory experiments carried out by
\citet{2009ApJ...696..706J} demonstrate that the temperature in the
condensation zone determines the formation pathways of carbonaceous
particles with a clear distinction between the resulting condensation
products. At temperatures higher than 3500~K, the resulting particles
are fullerene compounds and fullerene-like carbon grains (even in the
presence of hydrogen); at temperatures lower than 1700~K, soot is
formed with a large mixture of PAHs as by-products. This
high-temperature mechanism may be relevant to explain the fullerenes
in the RCBs \citep{Garcia-Hernandez:RcrB,2011arXiv1106.0563C}, but a
value of 3500~K is certainly not a typical temperature in the
circumstellar environment of carbon stars. It remains to be seen
whether such a formation route is also important for PNe.

Finally, a promising route in the astronomical settings presented
here, is that fullerenes form following the destruction of PAHs.
\citet{Micelotta:hotgas,Micelotta:shocks} studied the effects of
shocks on interstellar PAHs, and found that the PAH structure is at
least significantly altered (if not destroyed) for shocks with
velocities between 75--150 km s$^{-1}$ and in the hot post-shock
gas. PAHs first lose their peripheral H-atoms, and then the resulting
carbon clusters could assemble into fullerenes on fairly short
timescales. This would certainly also be a viable route in
PNe. Assembling fullerenes from PAH destruction could also explain the
different spatial distribution of PAHs and fullerenes in NGC~7023 (see
Fig.~\ref{Fig:NGC7023:C60}).

\section{Conclusions \& outlook}

The fullerene species C$_{60}$ and C$_{70}$ have been identified in a
variety of astronomical objects. It seems clear that fullerenes form
in the circumstellar environments of evolved stars, and that they
survive and thrive in the interstellar medium as well. It is less
clear precisely how they form, and what their excitation mechanism is,
and those topics remain a matter of debate and further research.  The
derived abundances indicate that $\sim$0.5\% of the cosmic carbon
could be in the form of C$_{60}$, and thus fullerenes represent a
significant component of interstellar matter. \\

Fullerenes and fullerene compounds have often been suggested as
possible carriers for some of the diffuse interstellar bands
(DIBs). Neutral C$_{60}$ is probably not a DIB carrier, but
C$_{60}^{+}$ has been suggested as a carrier for some DIBs
\citep{Foing:C60_1,Foing:C60_2,2009ApJ...700.1988M}.

Fullerene compounds offer many more possibilities. Fullerenes are
readily protonated \citep[see e.g. the review on C$_{60}$ cation
chemistry by][and references therein]{Bohme2009}, and it is thus quite
possible that fulleranes (hydrogenated fullerenes) might be present in
the ISM or in circumstellar environments as well. It has been
suggested that fulleranes could play a role in the formation of H$_2$
\citep{2009MNRAS.400..291C} and in interstellar
extinction. C$_{60}^{+}$ was found to be generally quite unreactive at
room temperature; nonetheless, many derivatization reactions have been
studied in laboratory experiments, including reactions with species
that have been detected in the ISM \citep[again, see][]{Bohme2009} and
could thus also be relevant for interstellar chemistry. Fullerenes are
also unique in their ability to lock up atoms and small molecules
inside their carbon cage. Such endohedral species have also been
proposed as DIB carrier candidates, but have only been poorly studied
in detail spectroscopically. Given the clear presence of fullerenes in
space, we encourage further laboratory work and theoretical
calculations on these species.

\begin{discussion}
  \discuss{Gu\'elin}{You were able to resolve spatially the PAH source
    from the C$_{60}$ source. Do you see any shift in position between
    the lines attributed to fullerenes?}  

  \discuss{Cami}{The spatial distribution shown is for the reflection
    nebula NGC~7023 as presented by \citet{2010ApJ...722L..54S}. In
    that source, most C$_{60}$ bands show contamination by
    neighbouring PAH features which certainly complicates matters; no
    shift in wavelength has been reported in that source.}

\discuss{Sternberg}{Does the preferential formation of PAHs versus
  C$_{60}$ in the hydrogen-rich environments depend on whether the
  hydrogen is ionized or not?}

\discuss{Cami}{The laboratory experiments I referred to used an
  atmosphere of Ar and H$_{2}$; I am not aware of experiments with
  protons. }

\discuss{Rawlings}{Would you care to comment on the expected degree of
  hydrogenation of the fullerenes in space?}

\discuss{Cami}{C$_{60}$ is quite easily protonated, so I would
  certainly expect to see hydrogenated fullerenes in space.}

\end{discussion}


\begin{thebibliography}{}

\bibitem[{Bakes} et~al.(2001){Bakes}, {Tielens}, {Bauschlicher}, {Hudgins}, \&
  {Allamandola}]{2001ApJ...560..261B}
{Bakes} E.L.O., {Tielens} A.G.G.M., {Bauschlicher}, Jr. C.W., {Hudgins} D.M.,
  {Allamandola} L.J., 2001, {\em ApJ\/} 560, 261

\bibitem[{Bettens} \& {Herbst}(1995){Bettens} \& {Herbst}]{1995IJMSI.149..321B}
{Bettens} R.P.A., {Herbst} E., 1995, {\em International Journal of Mass
  Spectrometry and Ion Processes\/} 149, 321

\bibitem[Bohme(2009)Bohme]{Bohme2009}
Bohme D.K., 2009, {\em Mass Spectrometry Reviews\/} 28, 672

\bibitem[{Cami} et~al.(2010){Cami}, {Bernard-Salas}, {Peeters}, \&
  {Malek}]{Cami:C60-Science}
{Cami} J., {Bernard-Salas} J., {Peeters} E., {Malek} S.E., 2010, {\em
  Science\/} 329, 1180

\bibitem[{Cataldo} \& {Iglesias-Groth}(2009){Cataldo} \&
  {Iglesias-Groth}]{2009MNRAS.400..291C}
{Cataldo} F., {Iglesias-Groth} S., 2009 400, 291

\bibitem[Choi et~al.(2000)Choi, Kertesz, \& Mihaly]{Choi_etal_2000}
Choi C.H., Kertesz M., Mihaly L., 2000, {\em The Journal of Physical Chemistry
  A\/} 104, 102

\bibitem[{Clayton} et~al.(2011){Clayton}, {De Marco}, {Whitney},
  et~al.]{2011arXiv1106.0563C}
{Clayton} G.C., {De Marco} O., {Whitney} B.A., et~al., 2011, {\em ArXiv
  e-prints\/}

\bibitem[{Clayton} et~al.(1995){Clayton}, {Kelly}, {Lacy},
  et~al.]{1995AJ....109.2096C}
{Clayton} G.C., {Kelly} D.M., {Lacy} J.H., et~al., 1995, {\em AJ\/} 109, 2096

\bibitem[{de Vries} et~al.(1993){de Vries}, {Reihs}, {Wendt},
  et~al.]{1993GeCoA..57..933D}
{de Vries} M.S., {Reihs} K., {Wendt} H.R., et~al., 1993, {\em Geochim.
  Cosmochim. Acta\/} 57, 933

\bibitem[Fabian(1996)Fabian]{PhysRevB.53.13864}
Fabian J., 1996, {\em Phys. Rev. B\/} 53, 13864

\bibitem[{Foing} \& {Ehrenfreund}(1994){Foing} \& {Ehrenfreund}]{Foing:C60_1}
{Foing} B.H., {Ehrenfreund} P., 1994, {\em Nat\/} 369, 296

\bibitem[{Foing} \& {Ehrenfreund}(1997){Foing} \& {Ehrenfreund}]{Foing:C60_2}
{Foing} B.H., {Ehrenfreund} P., 1997, {\em A\&A\/} 317, L59

\bibitem[{Frum} et~al.(1991){Frum}, {Engleman}, {Hedderich}, et~al.]{Frum:C60}
{Frum} C.I., {Engleman} R., {Hedderich} H.G., et~al., 1991, {\em Chem. Phys.
  Lett.\/} 176, 504

\bibitem[{Fulara} et~al.(1993){Fulara}, {Jakobi}, \&
  {Maier}]{1993CPL...211..227F}
{Fulara} J., {Jakobi} M., {Maier} J.P., 1993, {\em Chem. Phys. Lett.\/} 211,
  227

\bibitem[{Garc{\'{\i}}a-Hern{\'a}ndez}
  et~al.(2011){Garc{\'{\i}}a-Hern{\'a}ndez}, {Kameswara Rao}, \&
  {Lambert}]{Garcia-Hernandez:RcrB}
{Garc{\'{\i}}a-Hern{\'a}ndez} D.A., {Kameswara Rao} N., {Lambert} D.L., 2011,
  {\em ApJ\/} 729, 126

\bibitem[{Garc{\'{\i}}a-Hern{\'a}ndez}
  et~al.(2010){Garc{\'{\i}}a-Hern{\'a}ndez}, {Manchado}, {Garc{\'{\i}}a-Lario},
  et~al.]{Garcia-Hernandez:PN}
{Garc{\'{\i}}a-Hern{\'a}ndez} D.A., {Manchado} A., {Garc{\'{\i}}a-Lario} P.,
  et~al., 2010, {\em ApJ Lett.\/} 724, L39

\bibitem[{Gielen} et~al.(2011){Gielen}, {Cami}, {Bouwman}, \&
  {Min}]{Gielen:C60p-AGB}
{Gielen} C.A., {Cami} J., {Bouwman} J., {Min} M., 2011, {\em submitted to
  A\&A\/}

\bibitem[{Grishko} et~al.(2001){Grishko}, {Tereszchuk}, {Duley}, \&
  {Bernath}]{2001ApJ...558L.129G}
{Grishko} V.I., {Tereszchuk} K., {Duley} W.W., {Bernath} P., 2001, {\em ApJ
  Lett.\/} 558, L129

\bibitem[{Herbig}(2000){Herbig}]{Herbig:C60}
{Herbig} G.H., 2000, {\em ApJ\/} 542, 334

\bibitem[{Hony} et~al.(2002){Hony}, {Waters}, \& {Tielens}]{Hony:MgS}
{Hony} S., {Waters} L.B.F.M., {Tielens} A.G.G.M., 2002, {\em A\&A\/} 390, 533

\bibitem[{Houck} et~al.(2004){Houck}, {Roellig}, {van Cleve},
  et~al.]{Houck:IRS}
{Houck} J.R., {Roellig} T.L., {van Cleve} J., et~al., 2004, {\em ApJS\/} 154,
  18

\bibitem[{Iglesias-Groth} et~al.(2011){Iglesias-Groth}, {Cataldo}, \&
  {Manchado}]{2011MNRAS.413..213I}
{Iglesias-Groth} S., {Cataldo} F., {Manchado} A., 2011 413, 213

\bibitem[{J{\"a}ger} et~al.(2009){J{\"a}ger}, {Huisken}, {Mutschke}, {Jansa},
  \& {Henning}]{2009ApJ...696..706J}
{J{\"a}ger} C., {Huisken} F., {Mutschke} H., {Jansa} I.L., {Henning} T., 2009,
  {\em ApJ\/} 696, 706

\bibitem[{Joblin} et~al.(2002){Joblin}, {Toublanc}, {Boissel}, \&
  {Tielens}]{Joblin:MC}
{Joblin} C., {Toublanc} D., {Boissel} P., {Tielens} A.G.G.M., 2002, {\em
  Molecular Physics\/} 100, 3595

\bibitem[{Kr{\"a}tschmer} et~al.(1990){Kr{\"a}tschmer}, {Lamb},
  {Fostiropoulos}, \& {Huffman}]{1990Natur.347..354K}
{Kr{\"a}tschmer} W., {Lamb} L.D., {Fostiropoulos} K., {Huffman} D.R., 1990,
  {\em Nat\/} 347, 354

\bibitem[{Kroto} et~al.(1985){Kroto}, {Heath}, {Obrien}, {Curl}, \&
  {Smalley}]{1985Natur.318..162K}
{Kroto} H.W., {Heath} J.R., {Obrien} S.C., {Curl} R.F., {Smalley} R.E., 1985,
  {\em Nat\/} 318, 162

\bibitem[{Kroto} \& {McKay}(1988){Kroto} \& {McKay}]{1988Natur.331..328K}
{Kroto} H.W., {McKay} K., 1988, {\em Nat\/} 331, 328

\bibitem[Martin et~al.(1993)Martin, Koller, \& Mihaly]{PhysRevB.47.14607}
Martin M.C., Koller D., Mihaly L., 1993, {\em Phys. Rev. B\/} 47, 14607

\bibitem[{Mendez} et~al.(1988){Mendez}, {Kudritzki}, {Herrero}, {Husfeld}, \&
  {Groth}]{1988A&A...190..113M}
{Mendez} R.H., {Kudritzki} R.P., {Herrero} A., {Husfeld} D., {Groth} H.G.,
  1988, {\em A\&A\/} 190, 113

\bibitem[{Micelotta} et~al.(2010a){Micelotta}, {Jones}, \&
  {Tielens}]{Micelotta:hotgas}
{Micelotta} E.R., {Jones} A.P., {Tielens} A.G.G.M., 2010a, {\em A\&A\/} 510,
  A37+

\bibitem[{Micelotta} et~al.(2010b){Micelotta}, {Jones}, \&
  {Tielens}]{Micelotta:shocks}
{Micelotta} E.R., {Jones} A.P., {Tielens} A.G.G.M., 2010b, {\em A\&A\/} 510,
  A36+

\bibitem[{Milanova} \& {Kholtygin}(2009){Milanova} \&
  {Kholtygin}]{2009AstL...35..518M}
{Milanova} Y.V., {Kholtygin} A.F., 2009, {\em Astronomy Letters\/} 35, 518

\bibitem[{Misawa} et~al.(2009){Misawa}, {Gandhi}, {Hida}, {Tamagawa}, \&
  {Yamaguchi}]{2009ApJ...700.1988M}
{Misawa} T., {Gandhi} P., {Hida} A., {Tamagawa} T., {Yamaguchi} T., 2009, {\em
  ApJ\/} 700, 1988

\bibitem[{Moutou} et~al.(1999){Moutou}, {Sellgren}, {Verstraete}, \&
  {L{\'e}ger}]{Moutou:C60}
{Moutou} C., {Sellgren} K., {Verstraete} L., {L{\'e}ger} A., 1999, {\em A\&A\/}
  347, 949

\bibitem[Nemes et~al.(1994)Nemes, Ram, Bernath, et~al.]{Nemes:C60C70_temp}
Nemes L., Ram R., Bernath P., et~al., 1994, {\em Chem. Phys. Lett.\/} 218, 295

\bibitem[{Rubin} et~al.(2011){Rubin}, {Simpson}, {O'Dell},
  et~al.]{2011MNRAS.410.1320R}
{Rubin} R.H., {Simpson} J.P., {O'Dell} C.R., et~al., 2011, {\em MNRAS\/} 410,
  1320

\bibitem[Schettino et~al.({2002})Schettino, Pagliai, \&
  Cardini]{ISI:000174398500021}
Schettino V., Pagliai M., Cardini G., {2002}, {\em {JOURNAL OF PHYSICAL
  CHEMISTRY A}\/} {106}, 1815

\bibitem[{Scott} et~al.(1997){Scott}, {Duley}, \& {Pinho}]{1997ApJ...489L.193S}
{Scott} A., {Duley} W.W., {Pinho} G.P., 1997, {\em ApJ Lett.\/} 489, L193+

\bibitem[{Sellgren} et~al.(2007){Sellgren}, {Uchida}, \&
  {Werner}]{2007ApJ...659.1338S}
{Sellgren} K., {Uchida} K.I., {Werner} M.W., 2007, {\em ApJ\/} 659, 1338

\bibitem[{Sellgren} et~al.(2010){Sellgren}, {Werner}, {Ingalls},
  et~al.]{2010ApJ...722L..54S}
{Sellgren} K., {Werner} M.W., {Ingalls} J.G., et~al., 2010, {\em ApJ Lett.\/}
  722, L54

\bibitem[{Snow} \& {Seab}(1989){Snow} \& {Seab}]{1989A&A...213..291S}
{Snow} T.P., {Seab} C.G., 1989, {\em A\&A\/} 213, 291

\bibitem[{Sogoshi} et~al.(2000){Sogoshi}, {Kato}, {Wakabayashi},
  et~al.]{ISI:000087900400035}
{Sogoshi} N., {Kato} Y., {Wakabayashi} T., et~al., 2000, {\em {Journal of
  Physical Chemistry A}\/} 104, 3733

\bibitem[{Somerville} \& {Bellis}(1989){Somerville} \&
  {Bellis}]{1989MNRAS.240P..41S}
{Somerville} W.B., {Bellis} J.G., 1989 240, 41P

\bibitem[{Speck} et~al.(2009){Speck}, {Corman}, {Wakeman}, {Wheeler}, \&
  {Thompson}]{Speck:SiC}
{Speck} A.K., {Corman} A.B., {Wakeman} K., {Wheeler} C.H., {Thompson} G., 2009,
  {\em ApJ\/} 691, 1202

\bibitem[Stratmann et~al.({1998})Stratmann, Scuseria, \&
  Frisch]{ISI:000074726600006}
Stratmann R., Scuseria G., Frisch M., {1998}, {\em {Journal of Raman
  Spectroscopy}\/} {29}, 483

\bibitem[{Tielens}(2005){Tielens}]{Tielens:book}
{Tielens} A.G.G.M., 2005, {The Physics and Chemistry of the Interstellar
  Medium}

\bibitem[von Czarnowski \& Meiwes-Broer(1995)von Czarnowski \&
  Meiwes-Broer]{vonCzarnowski1995321}
von Czarnowski A., Meiwes-Broer K.H., 1995, {\em Chemical Physics Letters\/}
  246, 321

\bibitem[{Wang} et~al.(1995){Wang}, {Lin}, {Mesleh},
  et~al.]{1995JMatR..10.1977W}
{Wang} X.K., {Lin} X.W., {Mesleh} M., et~al., 1995, {\em Journal of Materials
  Research\/} 10, 1977

\bibitem[{Werner} et~al.(2004){Werner}, {Roellig}, {Low},
  et~al.]{Werner:spitzer}
{Werner} M.W., {Roellig} T.L., {Low} F.J., et~al., 2004, {\em ApJS\/} 154, 1

\bibitem[{Williams} et~al.(2008){Williams}, {Jenkins}, {Baldwin},
  et~al.]{2008ApJ...677.1100W}
{Williams} R., {Jenkins} E.B., {Baldwin} J.A., et~al., 2008, {\em ApJ\/} 677,
  1100

\bibitem[{Zhang} \& {Kwok}(2011){Zhang} \& {Kwok}]{ZhangKwok:proto-PNC60}
{Zhang} Y., {Kwok} S., 2011, {\em ApJ\/} 730, 126

\end{thebibliography}

\end{document}